\documentclass[12pt]{article}
\usepackage{amsmath,amssymb,amsthm,amsxtra,overpic,bbm,bm,epsfig,subfigure}
\usepackage[colorlinks=true,citecolor=blue,linkcolor=blue]{hyperref}
\usepackage{mathrsfs}
\usepackage{enumitem}
\usepackage{graphicx}
\usepackage{color}
\usepackage{comment}
\usepackage{epstopdf}
\usepackage{float}
\usepackage{cite}
\usepackage{multirow}
\usepackage{slashed,stmaryrd}

\def\thefootnote{\fnsymbol{footnote}}

\def\tabnotefont{\fontsize{9}{10}\selectfont}%
\newenvironment{tabnote}{\par\tabnotefont}{\par}

\begin{document}

\vspace{0.2cm}

\begin{center}
{\Large\bf Neutral-current background induced by atmospheric neutrinos at large liquid-scintillator detectors:}\\
{\Large\bf \underline{II. {Methodology for {\it in situ} measurements}}} 
\end{center}

\vspace{0.2cm}

\begin{center}
{\bf Jie Cheng$^{a}$}\footnote{Email: chengjie@ihep.ac.cn},
\quad
{\bf Yu-Feng Li$^{a, b}$}\footnote{Email: liyufeng@ihep.ac.cn},
\quad
{\bf Hao-Qi Lu$^{a}$}\footnote{Email: luhq@ihep.ac.cn},
\quad
{\bf Liang-Jian Wen$^{a}$}\footnote{Email: wenlj@ihep.ac.cn}
\\
\vspace{0.2cm}
{\small
$^a$Institute of High Energy Physics, Chinese Academy of Sciences, Beijing 100049, China\\
$^b$School of Physical Sciences, University of Chinese Academy of Sciences, Beijing 100049, China\\}
\end{center}

\vspace{1.5cm}

\begin{abstract}
Future large liquid-scintillator (LS) detectors are competitive with and complementary to the water-Cherenkov detectors on the searches for diffuse supernova neutrino background and nucleon decay. In a companion paper, we have performed a systematic calculation of the neutral-current (NC) background induced by atmospheric neutrino interactions on $^{12}{\rm C}$ nuclei in LS detectors, which are expected to be crucially important for the experimental searches for the diffuse supernova neutrino background and nucleon decay. In this paper, we perform a systematic study on the  measurement of the NC background and evaluate the associated uncertainties. We first exploit the characteristics of the NC background, in particular, the multiplicities of neutrons and pions, and the possible association with unstable residual nuclei. It turns out that the neutron multiplicity distribution is very powerful to discriminate among different models. Then, we develop a maximum-likelihood method to allow an {\it in situ} measurement of the NC interactions with a triple-coincidence signature. Finally, a data-driven approach is proposed to evaluate the uncertainty of the NC background in the search for the diffuse supernova neutrino background. We conclude that future large LS experiments like JUNO (Jiangmen Underground Neutrino Observatory) will be able to make a unique contribution to the worldwide data set to improve the prediction of atmospheric neutrino NC interactions on $^{12}$C.
\end{abstract}


\def\thefootnote{\arabic{footnote}}
\setcounter{footnote}{0}

\newpage

\section{Introduction}

Atmospheric neutrino observations have played an important role in the field of neutrino physics~\cite{Tanabashi:2018oca}, starting from the discovery of neutrino oscillations~\cite{Fukuda:1998mi,Kajita:2016cak} to the precision measurements of neutrino masses and mixing parameters~\cite{Abe:2017aap,Adamson:2013whj,Aartsen:2017nmd}. In the future, atmospheric neutrinos offer the chance to measure the unknown parameters of neutrino mass ordering~\cite{TheIceCube-Gen2:2016cap,Adrian-Martinez:2016fdl} and leptonic CP violation~\cite{Razzaque:2014vba,Kelly:2019itm}. Meanwhile, atmospheric neutrinos will also contribute to the irreducible backgrounds for rare event searches in large neutrino detectors, e.g., the diffuse supernova neutrino background (DSNB), and nucleon decay.

The DSNB is the integrated supernova (SN) neutrino flux from all past core-collapse events in the visible Universe, carrying very precise information on the cosmic star-formation rate, the average SN neutrino energy spectrum, and the rate of failed SNe~\cite{Ando:2004hc,Beacom:2010kk,Lunardini:2010ab}. It is worthwhile to mention that there is quite a challenge to extract this cosmological evolution information from future measurements but still conceivable to learn more via DSNB.  The existing and future large water-Cherenkov (wCh) and liquid-scintillator (LS) detectors, such as Super-K~\cite{Beacom:2003nk,Simpson:2018snj}, Hyper-K~\cite{Abe:2018uyc}, JUNO~\cite{An:2015jdp} and Theia~\cite{Askins:2019oqj}, have good potential to observe the DSNB via the inverse-beta-decay (IBD) reaction, $\overline{\nu}^{}_e + p \to e^+ + n$, which consists of a prompt signal of positron and a delayed signal of neutron capture. The LS detectors have intrinsically high efficiency of IBD detection due to excellent neutron tagging. For wCh detectors, Super-K has achieved to add dissolved gadolinium sulfate to feature greatly improved for neutron tagging, which can provide significant reference for the plan of adding gadolinium to water in Hyper-K.
However, both wCh and LS detectors suffer from the most critical background created by neutral-current (NC) interactions of atmospheric neutrinos with $^{16}$O and $^{12}$C, respectively.

The prediction of NC interactions induced by atmospheric neutrinos ($\nu^{}_{\rm atm}$) in either LS or water has large systematic uncertainties, due to large variations of model predictions and limited constraints from experimental data. The uncertainties come from the atmospheric neutrino flux, the NC interaction cross-section and the nucleus deexcitation processes. In the search for extraterrestrial $\bar\nu^{}_e$'s at KamLAND~\cite{Collaboration:2011jza}, the calculated NC background with the prompt energy between 7.5 MeV and 30 MeV has an uncertainty of 29\%. In the recent measurement of the neutrino-oxygen NC quasi-elastic (NCQE) cross section using atmospheric neutrinos at Super-K~\cite{Wan:2019xnl}, the uncertainties on flux, $\nu/\bar\nu$ ratio and cross sections for the NC processes other than NCQE are estimated to be 18\%, 5\% and 18\%, respectively. In a companion paper, referred to hereafter as ``the preceding paper"~\cite{Cheng:2020aaw}, we have performed a systematic study of $\nu^{}_{\rm atm}$-$^{12}{\rm C}$ NC interactions in LS. The rates and spectra of the NC backgrounds in LS are obtained by a two-fold calculation approach: the sophisticated generators \texttt{GENIE} and \texttt{NuWro} are used to calculate the neutrino-carbon interactions, then the \texttt{TALYS} package is used to predict the deexcitation processes of the residual nuclei. From these simulations we conclude there is a large uncertainty on the prediction of $\nu^{}_{\rm atm}$-$^{12}{\rm C}$ NC background for the DSNB search, i.e., 20\%, that originates from the variations of different nuclear models.

Reducing the uncertainty of the $\nu^{}_{\rm atm}$-$^{12}{\rm C}$ NC background prediction is of prominent importance for the searches for the DSNB. Motivated by this, we perform a systematic analysis of the calculated NC background from Ref.~\cite{Cheng:2020aaw} for the DSNB search. First we exploit the characteristics of the NC interactions, in particular, the correlations among the neutron multiplicity, the daughter residual nuclei and the prompt energy deposit of the particles (e.g., $p$, $\alpha$, neutrons and $\gamma$-rays) in the exclusive channels of the neutrino-$^{12}$C NC interactions. Such correlations can be utilized to measure {\it in situ} the NC background. Second, in a realistic detector, possible background sources that can mimic the signatures of NC interactions should be evaluated, because they may degrade the precision of the {\it in situ} measurement of the NC background. To be concrete, we choose the JUNO detector for a demonstration, and toy Monte Carlo datasets are built and the event selection criteria are developed.
Using the toy datasets, we employ a maximum likelihood method to extract the NC background with the triple-coincidence signature.
The uncertainty of the measured NC background is propagated to the DSNB signal region. Such a data-driven approach is promising to significantly reduce the uncertainty of the NC background prediction for the DSNB search in LS detectors, from around 20\% to 10\% level.

The remaining part of the paper is organized as follows. In Sec.~\ref{sec:atmNuC}, we highlight the key characteristics from the simulations of the $\nu^{}_{\rm atm}$-$^{12}{\rm C}$ NC interactions. Sec.~\ref{sec:InsituMeas} introduces the methodology for the {\it in situ} measurement of the NC interactions and demonstrates the method of reducing the uncertainty of NC background prediction for the DSNB search. Finally, we summarize our studies and conclude in Sec.~\ref{sec:summary}.

\section{Characteristics of $\nu^{}_{\rm atm}$-$^{12}{\rm C}$ NC interactions}
\label{sec:atmNuC}

In Ref.~\cite{Cheng:2020aaw}, we have calculated the NC background induced by atmospheric neutrino interactions with the $^{12}$C nuclei in the LS detectors. The Monte Carlo setup and simulated data sample used in this work are inherited from the preceding paper. In that calculation, the up-to-date fluxes of atmospheric neutrinos at the JUNO site provided by the Honda group~\cite{Hondahomepage} are used. Six representative nuclear models from the generators \texttt{GENIE (2.12.0)}~\cite{genie} and \texttt{NuWro (17.10)}~\cite{nuwro} are used to calculate the neutrino-nucleus interactions: one model from \texttt{GENIE} (i.e., Model-G) and five models from \texttt{NuWro} (i.e., Model-N$i$ for $i$=1,2,3,4,5). The notations and detailed descriptions of the six models are in the preceding paper~\cite{Cheng:2020aaw}, and a brief summary is presented in the following. The adopted models differ in the input values of the axial mass  $M^{}_{\rm A}$ in the parametrization of the nucleon axial-vector form factor. \texttt{GENIE} uses $M^{}_{\rm A}=0.99$ GeV~\cite{Kitagaki:1990vs} as the default setting. In \texttt{NuWro}, three different values, $M^{}_{\rm A}=$ 0.99, 1.35~\cite{AguilarArevalo:2010zc}, and 1.03~\cite{Bernard:2001rs} GeV are taken, respectively.Regarding the models of nuclear structure, \texttt{GENIE} uses the relativistic Fermi gas (RFG) model. In contrast, \texttt{NuWro} includes both RFG and the spectral function (SF) approach, which is used in Model-N(1-4) and Model-N5, respectively. Furthermore, to illustrate the two-body current effects in quasi-elastic scattering (QEL), the transverse enhancement model (TEM)~\cite{Bodek:2011ps} of the meson exchange current from \texttt{NuWro} is considered. It should be emphasized that only the axial mass $M^{}_{\rm A}$ in the treatment of QEL in \texttt{NuWro} has been changed. For both generators, we have employed their default setting for all other processes.

\begin{figure}[!h]
\centering
\includegraphics[width=8.5cm]{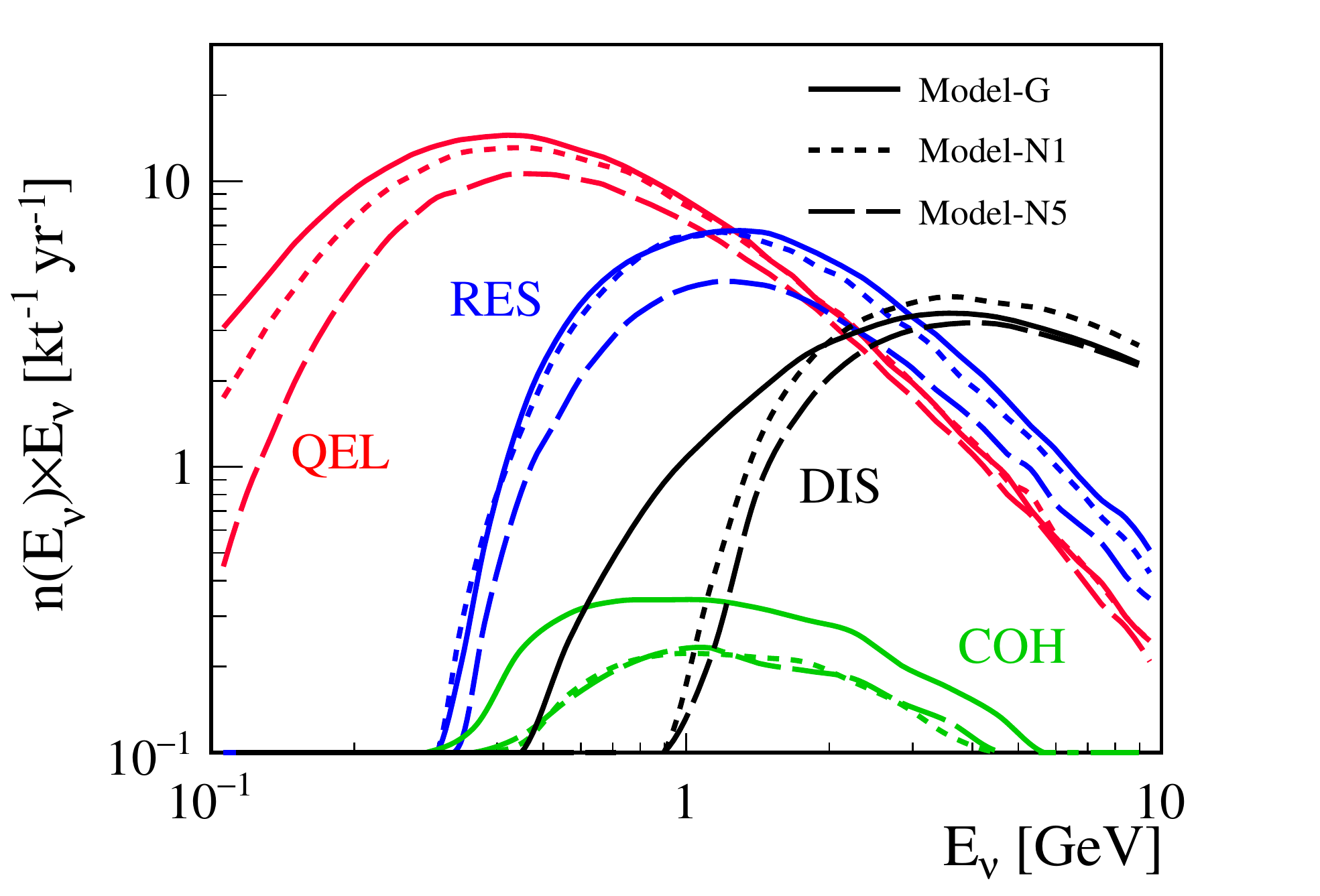}
\includegraphics[width=8.5cm]{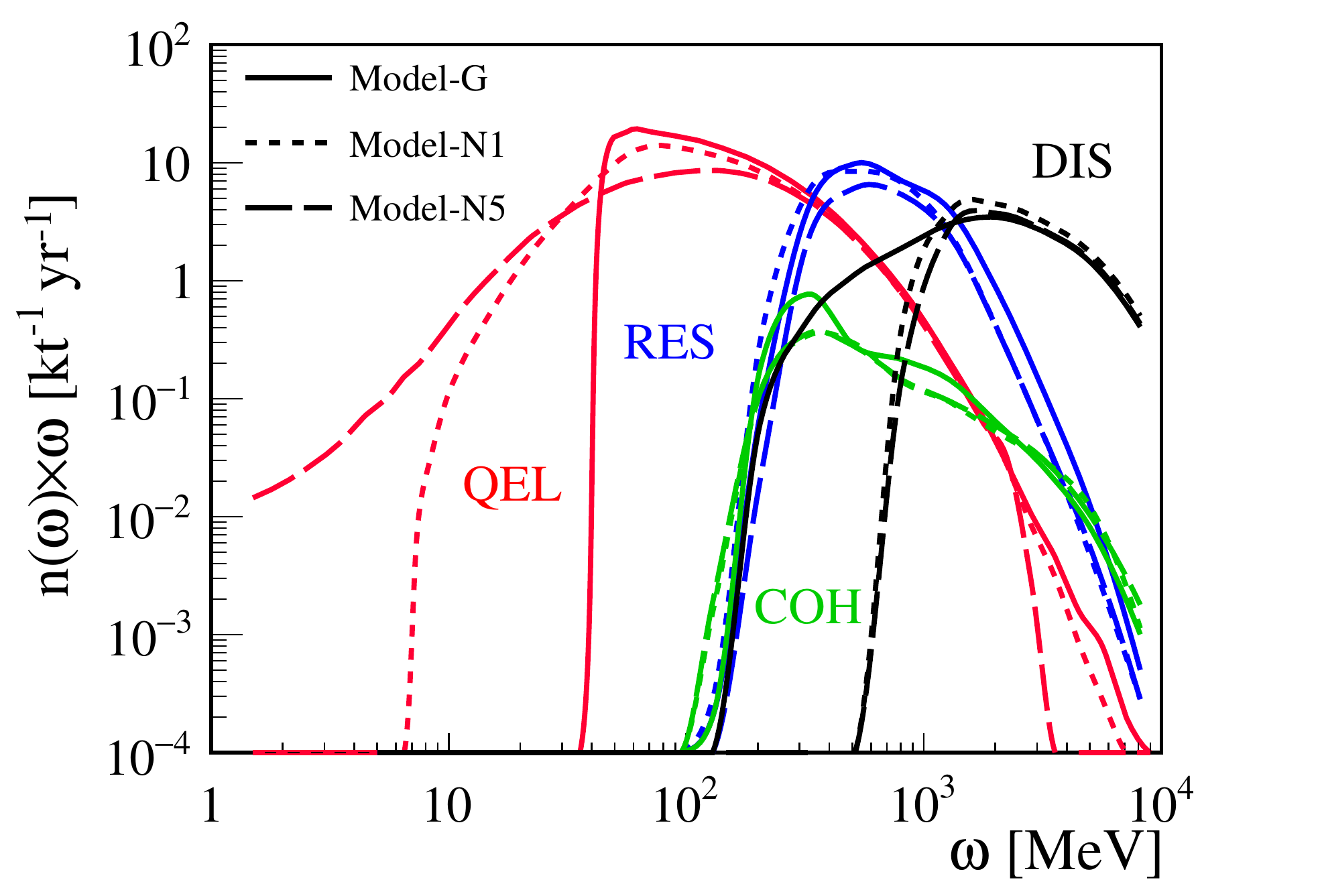}
\caption{Event rates of the QEL, RES, COH and DIS processes of neutrino-$^{12}$C NC interactions with respect to the incoming neutrino energy (left panel) and the energy transfer (right panel). The rates are obtained in the preceding paper~\cite{Cheng:2020aaw}, and multiplied by the neutrino energy ($E^{}_\nu$) and the energy transfer ($E^{}_{\rm trans}$) in two panels, respectively. Note that the event rates for all processes in the series of models (i.e., Model-N$i$ for $i$=1,2,3,4) are quite similar, thus only Model-G, Model-N1 and Model-N5 are shown. }
\label{fig:eNueTR}
\end{figure}

In different energy ranges from 100 MeV to GeV or even higher, the dominant contributions to the cross section comes roughly from QEL, coherent and diffractive production (COH), nuclear resonance production (RES), and deep inelastic scattering (DIS). The event rates of the QEL, RES, COH and DIS processes of the NC interactions of atmospheric neutrinos with $^{12}$C nuclei are shown in Fig.~\ref{fig:eNueTR}. The rates are displayed with respect to the neutrino energy and the energy transfer, respectively, in the left and right panel. The energy transfer ($\rm \omega \equiv E_{\nu} - E^{'}_{\nu}$) is defined as the energy difference between incoming and outgoing neutrinos. Since the predictions from Model-N$i$ for $i=1,2,3,4$ are quite similar, only Model-G, Model-N1 and Model-N5 are presented for illustration.

The deexcitation processes of the final-state nuclei produced in the NC interactions are handled by using \texttt{TALYS (1.8)}~\cite{talys}. Finally, a \texttt{GEANT4}-based Monte Carlo simulation is used to convert the kinetic energies of final-state particles in the NC interactions to the visible energies of the final events in the LS detectors. For simplicity, neither optical simulation nor specific detector geometry is involved, and only the quenching effect in LS is considered using the Birks' constants described in Ref.~\cite{An:2015jdp}. A summary of some observations from the preceding paper is helpful and relevant to this work. The QEL process of neutrino-$^{12}{\rm C}$ interactions is the predominant background for the DSNB search. The neutron multiplicity distribution in the final-state will be useful to scrutinize the nuclear models, e.g., the \texttt{GENIE} generator produces significantly higher event rates in the channels with more than two neutrons. The NC interactions with one neutron production may mimic the DSNB-like signal. The event rate for the exclusive processes with one neutron production is calculated to be $(16.5\pm 2.8)~{\rm kt}^{-1}~{\rm yr}^{-1}$ in the whole range of visible energies. The rate reduces to $(3.1\pm 0.5)~{\rm kt}^{-1}~{\rm yr}^{-1}$ if restricting into the energy window $11~{\rm MeV} \lesssim E^{}_{\rm vis} \lesssim 30~{\rm MeV}$ of interest. The associated uncertainty is about 20\%, representing the model variations of neutrino interactions. If adding in quadrature the extra uncertainty of 15\% from the calculations of atmospheric neutrino fluxes, we obtain the overall uncertainty will be 25\%.

Based on the signatures of DSNB, it requires a dedicated analysis of the NC background induced by atmospheric neutrinos, particularly a data-driven approach to utilize future experimental data to evaluate the nuclear models. In the following, for the DSNB search, we perform a systematic analysis of the most relevant NC background, i.e., the QEL processes.

\subsection{Characteristics of QEL interactions}

\subsubsection{Neutron multiplicity}
\label{sec:neuMulti}

In the QEL process of the neutrino-$^{12}$C NC interactions in the LS detectors, one or more nucleons may be knocked out from the carbon nucleus. In Ref.~\cite{Cheng:2020aaw}, the event rates for the NC interactions in the exclusive channels have been obtained and categorized by the associated neutron multiplicities, defined as the total numbers of produced neutrons from neutrino interaction and the subsequent deexcitation.
Moreover, many residual nuclei from the NC interactions are unstable isotopes. If the half-life is in a proper time window, the correlation between the isotopic decay and the parent NC interaction can be identified in LS. For this purpose, the NC interactions from QEL process can be divided into two categories:
\begin{itemize}
  \item Category I: the interactions associated with a suitably long-lived residual nucleus, i.e., $^{11}$C ($T^{}_{1/2}=20.39$ min, decay energy $Q^{}_{\beta^{+}\gamma}=1.98$ MeV), $^{10}$C ($T^{}_{1/2}=19.3$ s, $Q^{}_{\beta^{+}\gamma}=3.65$ MeV) and $^8$Li ($T^{}_{1/2}=0.840$ s, $Q^{}_{\beta^{-}}=16.0$ MeV).
  \item Category II: the other interactions. Their associated residual nuclei include stable isotopes, i.e., $^{10}$B, $^{9}$B, $^9$Be and $^6$Li; very short-lived isotopes, i.e., $^8$Be ($T^{}_{1/2}\sim6.7\times10^{-17}$ s) and $^9$B ($T^{}_{1/2}\sim8\times10^{-19}$ s); and very long-lived isotopes, i.e., $^{10}$Be ($T^{}_{1/2}=1.51\times10^6$ yr, $Q^{}_{\beta^-}=0.556$ MeV), $^{7}$Be ($T^{}_{1/2}=53.2$ d, $Q^{}_{\rm EC}=0.862$ MeV).
\end{itemize}

\begin{figure}[!t]
\centering
\includegraphics[width=7.5cm]{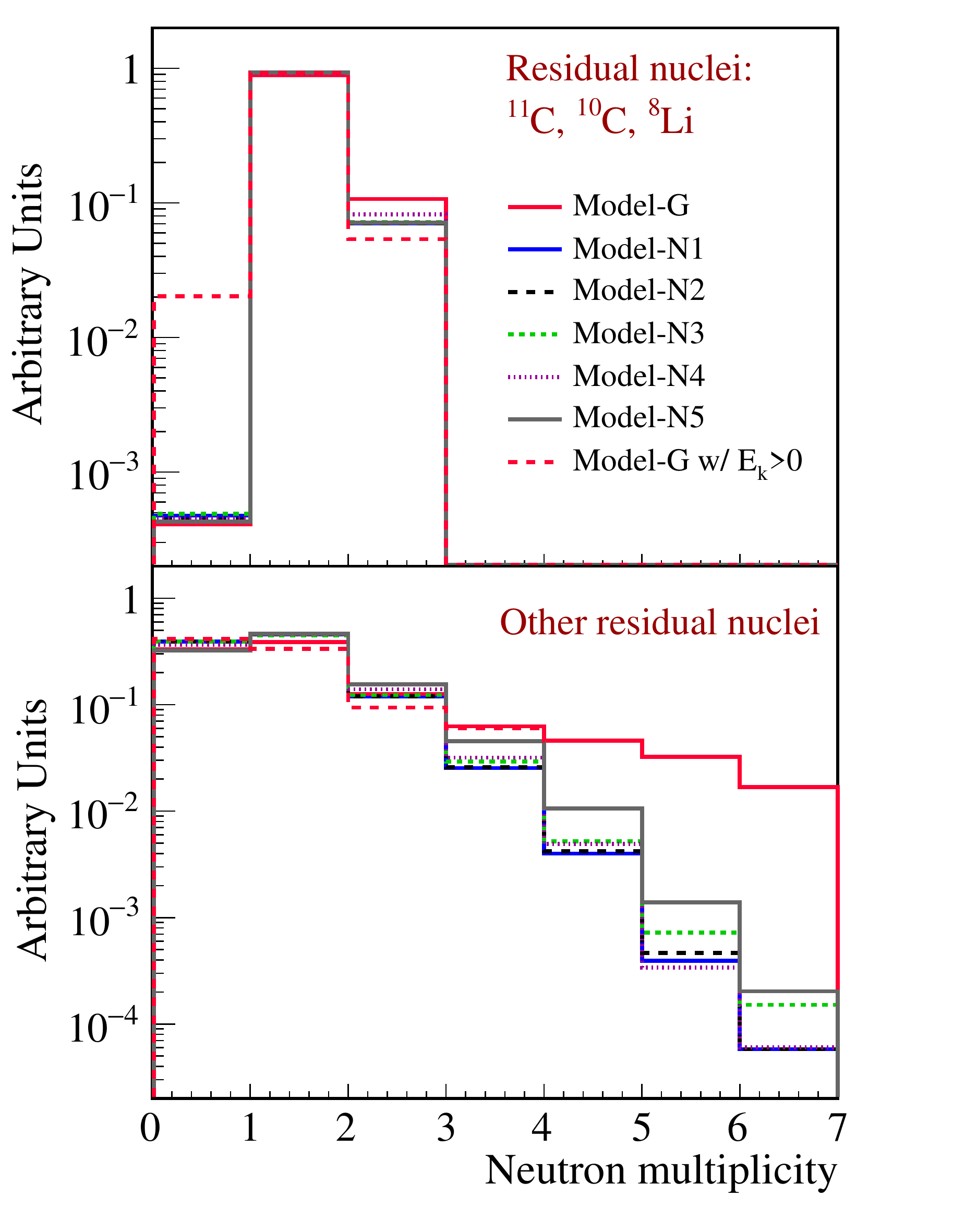}
\caption{Neutron multiplicity distributions for the QEL process of the NC interactions of atmospheric neutrinos with $^{12}$C. The top panel is for the final-states associated with $^{11}$C, $^{10}$C and $^{8}$Li, which have suitably long-lived half-lives, and the bottom panel is for the final-states with other residual nuclei, respectively. The red dashed line represents the \texttt{GENIE} model, for which neutrons with zero kinetic energies are removed.}
\label{fig:neuMulti}
\end{figure}

Fig.~\ref{fig:neuMulti} shows the neutron multiplicity distributions for the above two categories. For all six nuclear models, about $\gtrsim$~99.9\% of the interactions in Category I have at least one neutron, while for Category II this probability is larger than $(60-67)$\%. The probabilities rapidly decrease as the neutron multiplicities increase. Among the interactions with at least one neutron, the \texttt{GENIE} prediction for Category II has about 23\% probability to produce more than two neutrons, much higher than any of the \texttt{NuWro} models in which such probability is $(5-9)$\%. This indicates that the neutron multiplicity distribution might discriminate the \texttt{GENIE} and \texttt{NuWro} models. Although, it doesn't validate the models that differ in other important ways but have similar neutron multiplicities. 
However, we find that \texttt{GENIE} produces neutrons with zero kinetic energies in the final states of $n+p+\rm{^{10}B}$ and $2n+\rm{^{10}C}$, while the other \texttt{NuWro} models do not.
The issue of neutrons with zero kinetic energies is a known problem\footnote{\texttt{GENIE (2.12.0)} had an empirical intranuclear model (i.e., INTRANUKE `hA') to simulate hadron absorption, followed by multi-nucleon knockout. 
However, sometimes there is not enough energy or some oddities with binding energy subtractions, and some nucleons are produced with 
zero kinetic energies.}~\cite{genieDis}, which provokes an excess of very low energy (down to zero) nucleons and typically results from some nucleon knockout events with high multiplicities. Note that the up-to-date \texttt{GENIE} versions are expected to improve the modeling of the processes of low-energy nucleon knockout. We have checked the newer version of \texttt{GENIE (3.0.2)} with the INTRANUKE `hA' model, in which the fraction of neutrons with zero kinetic energies is smaller. However, the validity of the neutron productions with zero kinetic energies still needs to be determined.
Although these final state neutrons with zero kinetic energy from \texttt{GENIE} model seem to be odd, we still consider them rather than ignoring them in our following studies.
Due to the uncertainty of the neutron productions with zero kinetic energies, we include the red dashed line in Fig.~\ref{fig:neuMulti}, to show the effect of removing the neutrons at rest. We find that the probability of interactions in Category I without neutrons increases to about 2\% if neutrons with zero kinetic energies are removed. 

Neutron tagging is of great importance for probing the $\nu^{}_{\rm atm}$ NC interactions. In Ref.~\cite{Wan:2019xnl}, the NCQE events were selected by the nuclear deexcitation gamma and the neutron capture signal on hydrogen, however, the neutron tagging efficiency in water was only ($4-22$)\%. Thus, the recently-begun SuperK-Gd project~\cite{Simpson:2018snj}, with 0.02\% (0.2\%) by mass of gadolinium-sulfate dissolved in the Super-K detector, is targeted to reach 50\% (90\%) neutron-capture on gadolinium. As the first step, the goal of SuperK-Gd is to load 0.02\% of ${\rm Gd_2 (SO_4)_3}$~\cite{Nakajima:2020nu}, which was achieved on August 17, 2020~\cite{SuperK:2020news}. 
Hyper-K will plan to add 0.2\% of ${\rm Gd_2 (SO_4)_3}$ to the water and provide excellent neutron tagging capabilities~\cite{Abe:2014oxa}.
For LS detectors like JUNO, the neutron efficiency is intrinsically high (better than 99\%). 

\subsubsection{Triple-coincidence signature}
\label{sec:tripleSig}

\begin{figure}[!t]
\centering
\includegraphics[width=10cm]{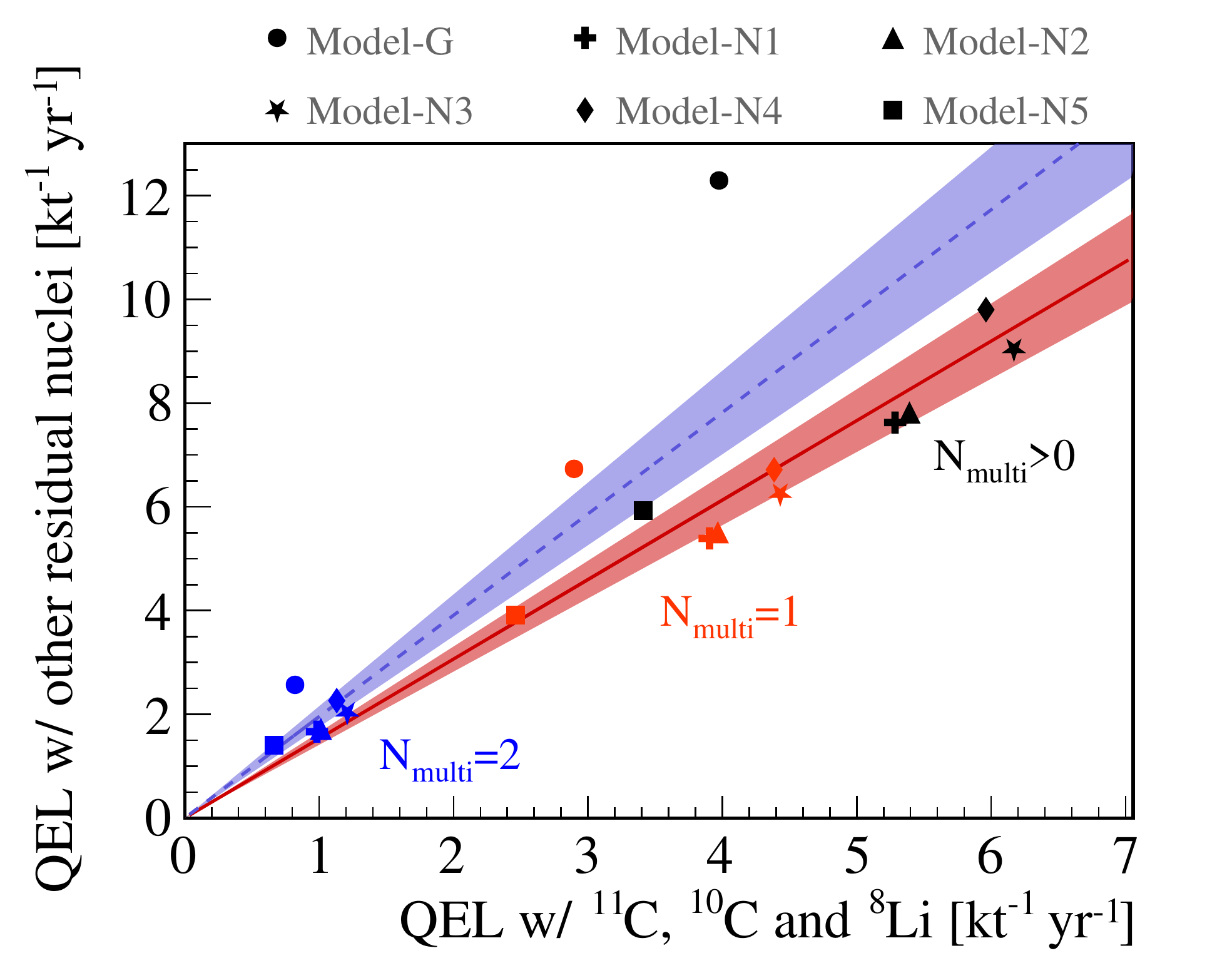}
\caption{Correlation between the rates of the NC interactions in the two categories. The different markers represent different nuclear models. The black points correspond to no selection on neutron multiplicity. The red and blue points represent the case of one tagged neutron ($N^{}_{\rm multi}=1$) and two tagged neutrons ($N^{}_{\rm multi}=2$), respectively. The solid and dashed lines are the linear fits to the points with $N^{}_{\rm multi}=1$ and $N^{}_{\rm multi}=2$, respectively, and the shaded bands represent the fitted error of the slopes.}
\label{fig:rate}
\end{figure}

\begin{figure}[!h]
\centering
\includegraphics[width=10cm]{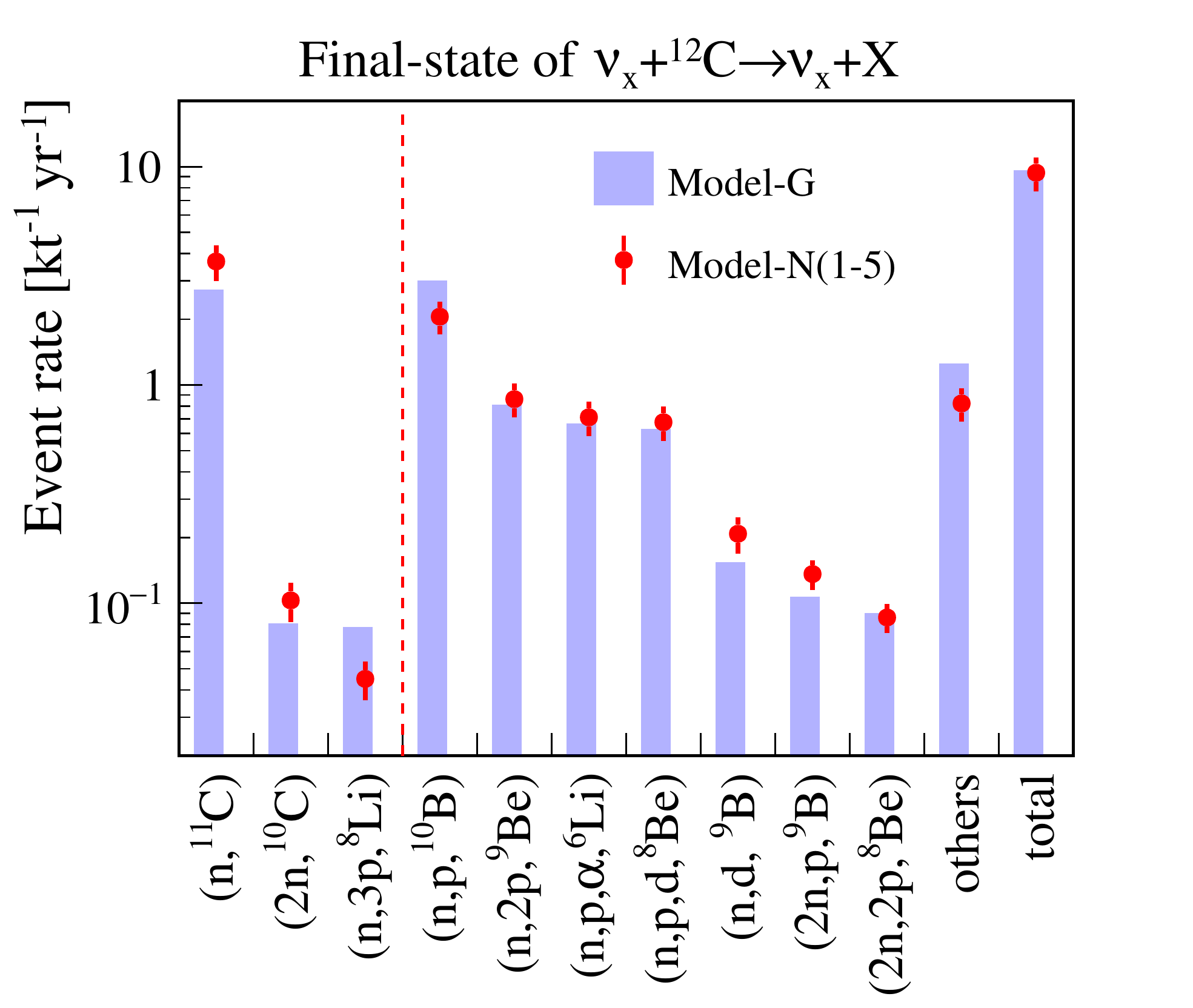}
\caption{Rates of the $\nu^{}_{\rm atm}$-$^{12}{\rm C}$ NC interactions in the exclusive channels associated with only one captured neutron and the visible prompt energy being less than 100 MeV. The channels are categorized by the final-states of the residual nuclei. The ``others" refers to the summation of the channels that each has a contribution less than 1\% and ``total" refers to the summation of all channels. The blue bars represent the predictions from Model-G.  The red circles with error bars stand for the mean value of the predictions from Model-N$i$ (for $i$=1,2,3,4,5) and the standard deviation. The dashed line separates the channels into two categories: triple-coincidences and double-coincidences.}
\label{fig:rate1n}
\end{figure}

The rates of the two categories have certain correlations among the six nuclear models, as shown in Fig.~\ref{fig:rate}. The values produced by \texttt{NuWro} models approximately have a linear dependency, whereas the value from \texttt{GENIE} is significantly off the trend line of the \texttt{NuWro} points.
If selecting the interactions with only one captured neutron, the values from all models show a slightly better linear-dependency, and a linear fit gives a slope of 1.53$\pm$0.12 (red solid line).
In the decay energy selection window of $^{11}$C and $^{10}$C, the contribution of $^8$Li is small. Hence if
not including $^8$Li, this slope will slightly increase to 1.55.
Furthermore, with a single tagged neutron, the signatures of Category I and II are triple-coincidence and double-coincidence, respectively, as shown in Fig.~\ref{fig:rate1n} (will be explained later). The latter typically consists of a prompt signal by fast-neutron recoil and the energy deposition of heavy charged particles ($p$, $d$ or $\alpha$) and a delayed signal by neutron capture on hydrogen, while the former has an additional signal from the unstable residual nucleus decaying at a later time. If selecting the interactions with two tagged neutrons, the linear dependency is 1.95$\pm$0.20 (blue dashed line). Such linear dependencies can be used to extrapolate the rate of one category from the other. In Sec.~\ref{sec:InsituMeas}, we will discuss how to measure the interactions of Category I. Moreover, as the existence of these neutrons with zero kinetic energy is doubtful, the slope of the red line changes to 1.48$\pm$0.06 if removing the neutrons with zero kinetic energy in Model-G, which indicates a more prominent linear dependency.
It should be noted that current model variations employed here are  limited. Other up-to-date neutrino event generators, such as \texttt{NUANCE}~\cite{Casper:2002sd}, \texttt{NEUT}~\cite{Hayato:2009zz} and \texttt{GiBUU}~\cite{Buss:2011mx} , will be added in future works to improve uncertainty evaluations.

\begin{figure}[!t]
\centering
\includegraphics[width=8cm]{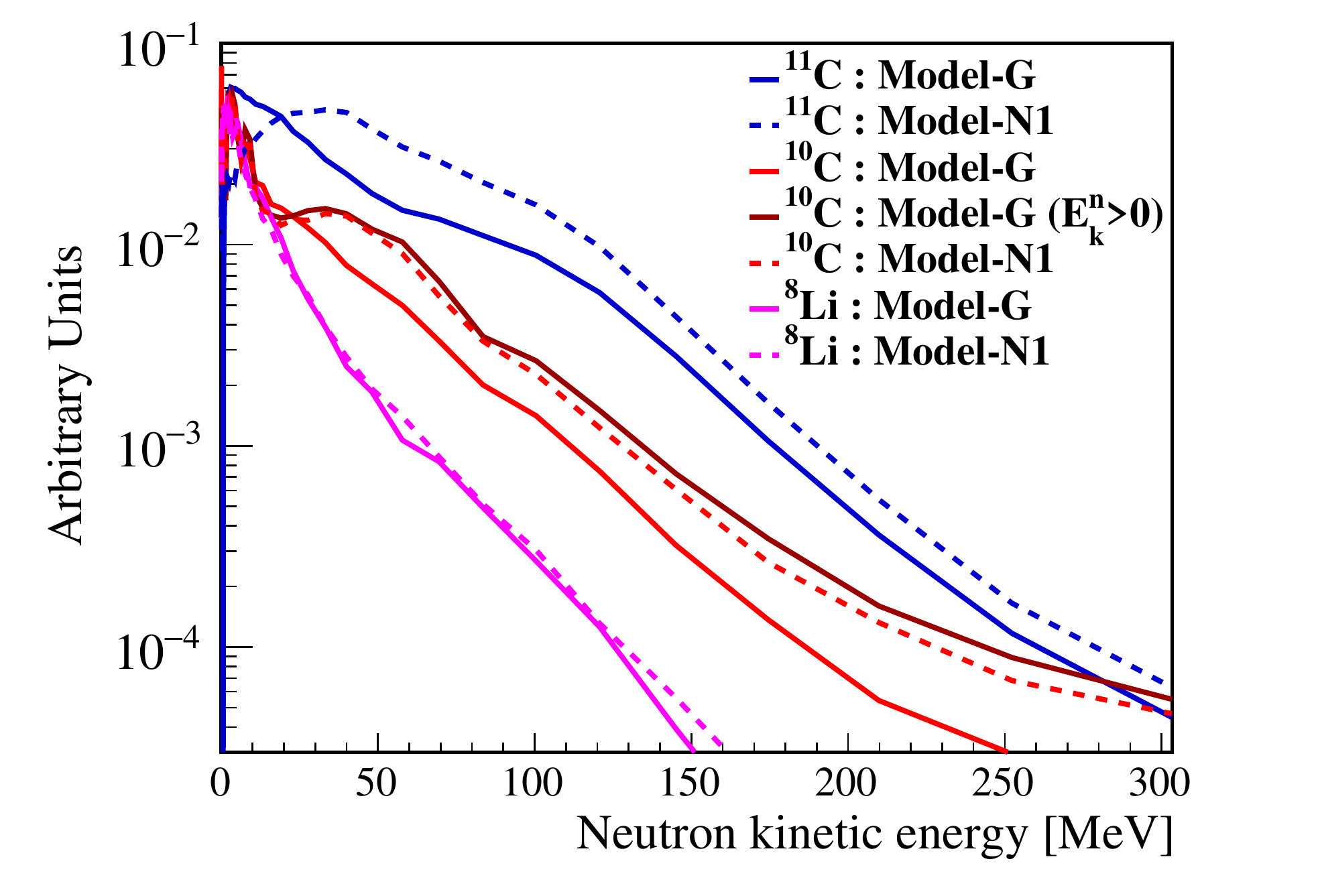}
\includegraphics[width=8cm]{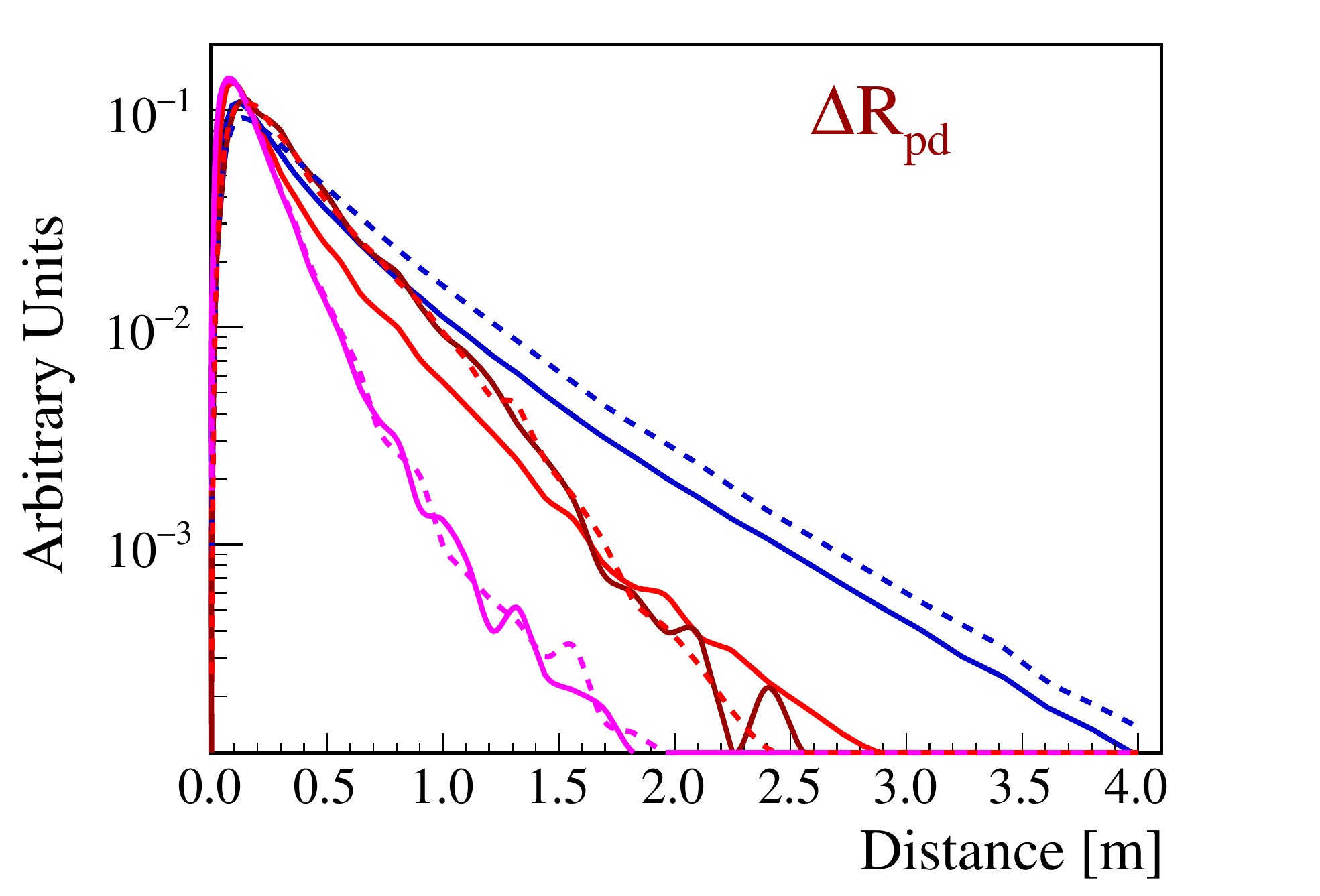}
\caption{Distributions of the kinetic energies of fast neutrons (left panel) and the distances between the fast neutron recoils (deposited-energy weighted vertexes) and the isotope decays (right panel) in the NC triple-coincidences associated with $^{11}$C, $^{10}$C and $^{8}$Li, respectively. The notation ``$\rm E^{n}_{k}>0$" refers to the requirement that the NC events should accompany neutrons with non-zero kinetic energies.}
\label{fig:distrR}
\end{figure}

Fig.~\ref{fig:rate1n} summarizes the final states and corresponding rates of the selected NC interactions associated with only one captured neutron
and the visible energy of the prompt signal being less than 100 MeV. The fractions of the channels in the series of models (i.e., Model-N$i$ for $i$=1,2,3,4) are quite similar, thus the average value and the standard deviation of the predictions from these five models are shown with error bars, and labeled by ``Model-N(1-5)". The exclusive channels are categorized into triple-coincidences and double-coincidences, which are separated by the dashed line. The triple-coincidences are dominated by the channels associated with $^{11}$C and $^{10}$C, while the double-coincidences have several major contributors. For both \texttt{GENIE} and \texttt{NuWro} models, the ratio of the total double-coincidences to the total triple-coincidences is about 3$/$2. By definition, both the triple-coincidence and the double-coincidence have only one tagged neutron. However, a few channels with double-neutron in the final states (i.e., $2n+{^{10}\textrm{C}}$, $2n+2p+{^{8}\textrm{Be}}$ and $2n+p+{^{9}\textrm{B}}$) fall into these categories, and it can be explained by two effects. First, the energetic neutrons may disappear due to their inelastic interactions with $^{12}$C. To qualitatively investigate this effect, we use \texttt{TALYS} to calculate the cross sections of the exclusive $n-^{12}{\rm C}$ reactions at different incident neutron energies. Then, taking the $2n+{^{10}\textrm{C}}$ final-state as an example, the neutron energy distribution in Fig.~\ref{fig:distrR} is used to calculate the integrated cross sections of the exclusive reaction channels. We find that about 7\% neutrons of the $2n+{^{10}\textrm{C}}$ final-state will vanish mainly via the  $^{12}{\rm C}(n, \alpha)^{9}{\rm Be}$ that is the abbreviation of $n + ^{12}\textrm{C}\rightarrow \alpha + ^{9}\textrm{Be}$ (4.8\%), $^{12}{\rm C}(n, d\alpha)^{7}{\rm Li}$ (0.6\%), $^{12}{\rm C}(n, d)^{11}{\rm B}$ (0.5\%), $^{12}{\rm C}(n, p)^{12}{\rm B}$ (0.5\%) and other sub-dominant processes. This simplified calculation qualitatively explains the existence of the final states with double-neutron in Fig.~\ref{fig:rate1n}. The rates in Fig.~\ref{fig:rate1n} are in fact obtained by a \texttt{GEANT4} simulation, which takes into account the neutron propagation.

To select a triple-coincidence signature, there are three pairs of time intervals and distances: ($\Delta t^{}_{\rm pn}$, $\Delta R^{}_{\rm pn}$) for the prompt and neutron-capture pair, ($\Delta t^{}_{\rm nd}$, $\Delta R^{}_{\rm nd}$) for the neutron-capture and isotopic decay pair, and ($\Delta t^{}_{\rm pd}$, $\Delta R^{}_{\rm pd}$) for the prompt and isotopic decay pair. Most neutrons from the NC interactions are fast neutrons, and the neutrons associated with $^{11}$C have higher kinetic energies (see the left panel of Fig.~\ref{fig:distrR}), resulting in a wider probability-density-function (PDF) of $\Delta R^{}_{\rm pd}$ for $^{11}$C than the PDFs for $^{10}$C and $^{8}$Li, as shown in the right panel of Fig.~\ref{fig:distrR}. For Model-G, if selecting the NC events with non-zero kinetic energy neutrons, the PDF for $^{10}$C (the dark red line) becomes close to that from Model-N1.
The PDF of $\Delta R^{}_{\rm pd}$ obtained from Monte Carlo simulations will be used for the fitting later. The $\Delta t^{}_{\rm pd}$ distribution follows $\sum B^{}_i\cdot e^{-t/\tau^{}_i}/\tau^{}_i$, where $B^{}_i$ and $\tau^{}_i$ are the production fractions and lifetimes of the residual isotopes.

The triple-coincidence signature will be used to measure the NC interactions in the following section. Similar triple-coincidence signatures have been widely used to study the cosmogenic long-lived isotopes in LS detectors, such as $^9$Li/$^8$He which is the most important background for reactor antineutrinos (e.g., see Refs.~\cite{An:2012eh,Adey:2018zwh}), and $^{10}$C which is a dominant background for solar neutrinos (e.g., see Refs.~\cite{Bellini:2013pxa,Abe:2009aa}).
In these analyses, the prompt signal consists of the energetic muon signals and the possible spallation neutron recoils. The spectral shapes of the isotopic decays and the times since the preceding muons were used to extract the production of cosmogenic isotopes.

\section{Methodology for {\it in situ} measurements of QEL interactions}
\label{sec:InsituMeas}

\subsection{Mock data set}

Based on the above characteristics, we develop a maximum-likelihood method to measure {\it in situ} the NC interactions with the triple-coincidence signature. To be concrete, we choose the JUNO detector as an example, as the JUNO detector is a suitable representative for future LS detectors. The mock data samples are produced, including $\nu^{}_{\rm atm}$-$^{12}{\rm C}$ NC interactions with a single neutron, DSNB, reactor antineutrinos, cosmogenic products ($^{11}$C/$^{10}$C, fast neutrons and $^9$Li/$^8$He), and accidental coincidences. The details of the JUNO detector are not required, instead, the rates are taken from the publicly available references. We concentrate on the demonstration of the method and leave the complex detector effects for future works by the experimental collaboration. Table.~\ref{tab:rateMC} summarizes the rates of individual MC samples used in this work, and some comments are helpful.
\begin{itemize}
  \item The initial rate and spectral shape of reactor antineutrinos are taken from Ref.~\cite{An:2015jdp}.
  \item The long-lived cosmogenic isotopes like $^{11}$C, $^{10}$C, $^8$Li and $^6$He are correlated with their preceding muons. However, after muon veto selection, the residual events can be regarded to be randomly distributed in time and space. They can coincide with the physical double-coincidence to form an accidental triple-coincidence, similar to the natural radioactivity. Many muon-induced isotopes have strong correlations with spallation neutrons, e.g., see Refs.~\cite{Zhao:2016brs,Abusleme:2020zyc}. Taking the veto strategy developed in Ref.~\cite{Zhao:2016brs}, the rates of $^{10}$C, $^8$Li and $^6$He can be suppressed by a factor of 310, 94 and 78, respectively, while maintaining a live-time efficiency of 84\%. Due to the long half-life of $^{11}$C, that veto strategy only reduces $^{11}$C approximately by 10\%. As also pointed out in Ref.~\cite{Zhao:2016brs}, the simulations by \texttt{FLUKA} and \texttt{GEANT4} predict different yields. In the recent solar neutrino study at JUNO~\cite{Abusleme:2020zyc}, an extrapolation based on the measurements from KamLAND~\cite{Abe:2009aa} and Borexino~\cite{Bellini:2013pxa} is used. All these calculations predict that $^{11}$C has the largest yield. Thus, in this paper we only consider the cosmogenic $^{11}$C and use the larger rate value 2300~${\rm kt}^{-1}~{\rm d}^{-1}$ from Ref.~\cite{Abusleme:2020zyc}, furthermore we quote the veto strategy and efficiency from Ref.~\cite{Zhao:2016brs}.
  \item The rate of the cosmogenic $^9$Li/$^8$He is taken from Ref.~\cite{An:2015jdp}, and the veto strategy and corresponding efficiency for $^9$Li/$^8$He is quoted from Ref.~\cite{Zhao:2016brs}. The fast-neutron rate is scaled from Ref.~\cite{An:2015jdp} according to the higher muon rate in Ref.~\cite{Abusleme:2020zyc}, and the fast-neutron energy spectrum is assumed to be flat.
  \item The energy spectra and spatial distributions of natural radioactivity is taken from Ref.~\cite{Abusleme:2020zyc}, where the $\alpha$'s are rejected by a pulse shape discriminator (PSD).
  \item The NC backgrounds induced by atmospheric neutrinos are taken from Fig.~\ref{fig:rate1n}. The average values of the \texttt{NuWro} models are used. Both the \texttt{GENIE} and \texttt{NuWro} values are tested for comparison.
  \item For the charged-current (CC) backgrounds induced by atmospheric neutrinos, it includes the CC interactions on $^{12}$C and protons (i.e., the IBD by the atmospheric $\overline\nu^{}_e$). The CC interactions on $^{12}$C are calculated using \texttt{GENIE} and \texttt{NuWro}. It shows that the prompt energy is larger than 100 MeV if a neutron-capture signal is required, thus the contribution of $\nu^{}_{\rm atm}$-$^{12}{\rm C}$ CC backgrounds is neglected in the mock data set. As for the IBD events induced by the atmospheric $\overline\nu^{}_e$, the rate is calculated with the up-to-date fluxes of atmospheric neutrinos at JUNO site provided by the Honda group~\cite{Hondahomepage} and the IBD cross section in Ref.~\cite{Strumia:2003zx}.
  \item The IBD rate from DSNB is calculated to be 0.30~${\rm kt}^{-1}~{\rm yr}^{-1}$ for the energy range from 7.5 MeV to 100 MeV, by using a DSNB model with the rate of core-collapse supernovae $R^{}_{\rm CCSN}=10^{-4}$~${\rm yr}^{-1}~{\rm Mpc}^{-3}$, the average energy of the core-collapse supernovae (CCSN) $\langle E\rangle=$14 MeV and the fraction of the failed supernovae rate $f^{}_{\rm BH}$=0.27. The parameters $R^{}_{\rm CCSN}$, $\langle E\rangle$ and $f^{}_{\rm BH}$ may have broad variations ~\cite{Priya:2017bmm}.
\end{itemize}
The mock data sets are produced based on Table.~\ref{tab:rateMC}. First, the individual MC samples are generated randomly in time and the LS volume, and the correlations inside the double-coincidences and triple-coincidences are automatically kept. Then, the independent data sets are combined and sorted by time to form the mock data samples. A set of mock data sets are produced assuming different experimental exposures.

\begin{table}[!ht]
\centering
\caption{Rates of the individual toy MC data samples in this work. }
\label{tab:rateMC}
\footnotesize
\begin{tabular}{cccc}
\hline
  &  Rate $^{\rm a}$  &  \multicolumn{2}{c}{Reduction} \\
\cline{3-4}
     &   ($\rm{kt}\cdot\rm{yr})^{-1}$   &   [$E,\Delta R,\Delta T$]  &  [PSD, $\mu$-veto]   \\
\hline
\multicolumn{4}{c}{\bf $\nu^{}_{\rm atm}$ NC Triple coincidence } \\
\hline
 & 2.74 ($^{11}$C) & 65.0\%  &  \multirow{3}{*}{95\% $\cdot$ 84\%}     \\
(GENIE) & 0.08 ($^{10}$C) & 62.5\%  &      \\
 & 0.08 ($^{8}$Li) & 10.0\%  &     \\
\\
 & 3.68 ($^{11}$C)  & 79.6\%  & \multirow{3}{*}{95\% $\cdot$ 84\%}   \\
(NuWro) & 0.10 ($^{10}$C)  & 80\%   &     \\
 &  0.05 ($^{8}$Li)  & 9.4\%  &     \\
\\
\multicolumn{4}{c}{\bf $\nu^{}_{\rm atm}$ NC Double coincidence} \\
\hline
(GENIE) &  6.73 & 70.0\% &  \multirow{2}{*}{95\% $\cdot$ 84\%}  \\
(NuWro) &  5.56 & 87.5\% &  \\
\\
\multicolumn{4}{c}{\bf Other Double coincidence}\\
\hline
$\nu^{}_{\rm atm}$ CC &  0.21 &   84.4\%  & 20\% $\cdot$ 84\%  \\
Fast-neutron   &  0.38 &   92.5\%  & 95\% $\cdot$ 84\%\\
DSNB    &      0.30    &   86.7\%  &   20\% $\cdot$ 84\% \\
Reactor $\bar\nu^{}_e$  &  1515  &  0.84\%  & 20\% $\cdot$ 84\% \\
$^{9}$Li/$^{8}$He    &  1533  &  14.7\%  & 20\% $\cdot$ 2\% \\
\\
\multicolumn{4}{c}{\bf Accidental coincidence} \\
\hline
   & Hz/kton & $\epsilon(E^{}_d)$ & N$^{}_{acc}$ $^{\rm a}$ \\
\hline
$^{11}$C from $\mu$ & 0.0267  &  100\% &  14.7 \\
Radioactivity  & 0.09 &  2.0\% &  1.0 \\
\hline
\end{tabular}
\begin{tabnote}
  $^{\rm a}$ Accidental background number in a decay selected window.\\
\end{tabnote}
\end{table}

Using the notations in Sec.~\ref{sec:tripleSig}, the following criteria are applied to the mock data set to select triple-coincidences: $E^{}_{\rm p}\in (7.5, 100.0)$ MeV, neutron multiplicity $N^{}_{\rm multi} = 1$, $E^{}_{\rm n}\in (1.8, 2.6)$ MeV,  $\Delta t^{}_{\rm pn}\in (1.0~{\rm \mu s}, 1.0~{\rm ms})$, $\Delta R^{}_{\rm pn}<1.5$ m, $E^{}_{\rm d}\in (1.0, 3.5)$ MeV, $\Delta t^{}_{\rm nd}\in (1.0~{\rm ms}, 4.0~{\rm hrs})$, $\Delta R^{}_{\rm pd}<2.2$ m,
where $E^{}_{\rm p}$, $E^{}_{\rm n}$ and $E^{}_{\rm d}$ are the energies of the prompt, neutron-like and decay-like signals. The prompt energy cut will remove all geo-neutrinos, a significant portion of the reactor antineutrinos, as well as all accidental coincidences with a natural radioactivity as the prompt. The $E^{}_{\rm d}$ cut covers the energy spectra of both $^{11}$C and $^{10}$C decays, and removes $\sim$88\% of $^8$Li. To further reduce the accidental triple-coincidences, a PSD can be constructed utilizing the scintillation time profile and applied to the prompt signal, and the detailed PSD study is in preparation and will be published elsewhere. In this paper,
we expect that the efficiency of the $\nu_{\rm atm}-^{12}$C NC interactions can achieve 95\%, while only 20\% of the $e$-like and $\gamma$-like prompt signals survive, which is a reasonable option according to the PSD study at JUNO~\cite{An:2015jdp}. The PSD optimization for specific detectors should take into account the detailed detector parameters, and it is nontrivial and beyond the scope of this work. The efficiencies due to the above selection criteria are shown in Table.~\ref{tab:rateMC}. Both the natural radioactivity and the cosmogenic $^{11}$C may mimic the decay-like signature of the NC triple-coincidence, and their raw rates and accidental coincidence rates are also listed in Table.~\ref{tab:rateMC}.
To calculate the accidental coincidences caused by the intrinsic $^{238}$U, $^{232}$Th, $^{40}$K and $^{210}$Pb radioactivity in LS, a `median' LS radio-purity level is considered ($10^{-16}\,\rm{g\,g}^{-1}$ $^{238}$U/$^{232}$Th, $10^{-17}\,\rm{g\,g}^{-1}$ $^{40}$K and $10^{-23}\,\rm{g\,g}^{-1}$ $^{210}$Pb), which is assumed to be 10 times worse than the `ideal' radiopurity in Ref.~\cite{An:2015jdp}. Under this assumption, the cosmogenic $^{11}$C will dominate the accidental triple-coincidences.

\subsection{Un-binned maximum-likelihood method}

\begin{figure}[!ht]
\centering
\includegraphics[width=8cm]{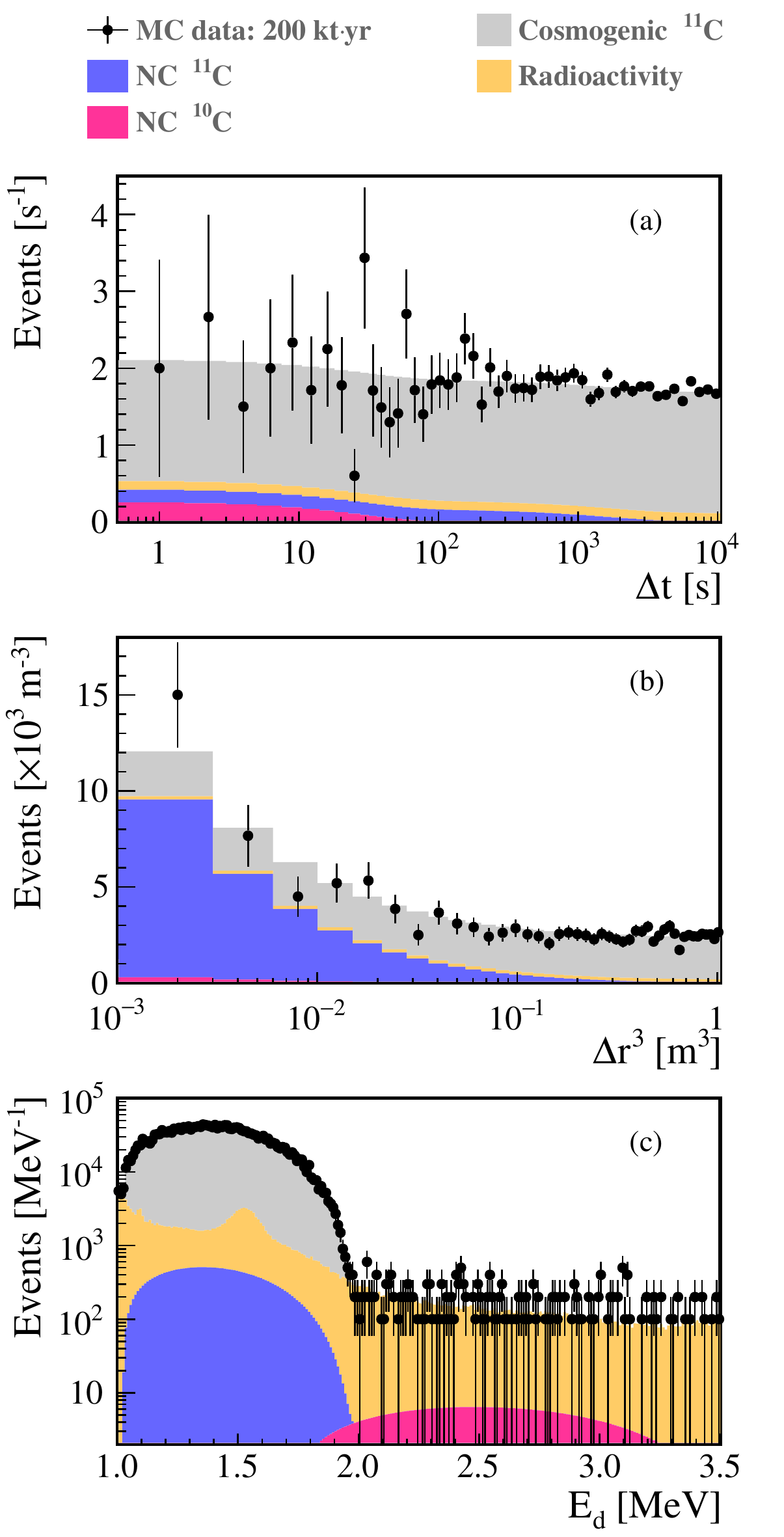}
\caption{Example of the distributions of the selected triple-coincidence candidates (black points) using the \texttt{GENIE} prediction with 200~${\rm kt}\cdot{\rm yr}$ exposure: the time interval between the prompt and third events (a), the cubic distance between the prompt and third events (b), and the energy of the third event (c). A combined fit to these distributions statistically distinguishes the two NC interactions associated $^{10}$C and $^{11}$C from the cosmogenic $^{11}$C and radioactivity.}
\label{fig:combFit}
\end{figure}

For the selected triple-coincidence candidates in each mock data set, the time interval ($\Delta t$) and the cubic-distance ($\Delta r^3$) between the prompt and third events, and the energy of the third event ($E^{}_{\rm d}$) are fit to the PDF of Eq.~(\ref{eqn:pdftot}), using an un-binned maximum likelihood (ML) method.
\begin{equation}\label{eqn:pdftot}
F(\Delta t, \Delta r^{3}, E^{}_{\rm d}) = \sum_i\frac{N^{}_i}{\tau^{}_i}\cdot e^{-\Delta t/\tau^{}_i} \cdot D^i_{\rm nc}(\Delta r^{3}) \cdot S^i_{\rm nc}(E^{}_{\rm d}) +  \frac{1}{T}\cdot D^{}_{\rm acc}(\Delta r^{3}) \cdot \sum_j N^{j}_{\rm acc}\cdot S^j_{\rm acc}(E^{}_{\rm d})
\end{equation}
The first term in Eq.~(\ref{eqn:pdftot}) represents the true triple-coincidences from NC interactions. The number of triple-coincidences associated with each isotope ($^{11}$C or $^{10}$C) is $N^{}_i$ with $\tau^{}_i$ as the decay lifetime, while the contribution from $^8$Li is too small to be included in the fitting. The spatial distribution and the $\beta$ energy spectrum of each isotope are denoted by $D^i_{\rm nc} (\Delta r^{3})$ and $S^i_{\rm nc}(E^{}_{\rm d})$, respectively. The second term in Eq.~(\ref{eqn:pdftot}) represents the accidental triple-coincidences with $T$ as the coincidence window, in which the $\Delta t$ has a flat distribution. $D^{}_{\rm acc} (\Delta r^{3})$ accounts for approximately flat contributions in space. The superscript $j$ represents the two main contributors: the cosmogenic $^{11}$C and the natural radioactivity. $N^{j}_{\rm acc}$  and $S^j_{\rm acc}(E^{}_{\rm d})$ account for the event numbers and energy spectra, respectively. The parameters $N^{}_i$ and $N^{j}_{\rm acc}$ are free, and $\tau^{}_i$ is fixed at 29.42 min and 27.78 s for $^{11}$C and $^{10}$C, respectively. $D^i_{\rm nc} (\Delta r^{3})$ is taken from the \texttt{GEANT4} simulation, as shown in the right panel of Fig.~\ref{fig:distrR}. $S^i_{\rm nc}(E^{}_{\rm d})$ and $S^j_{\rm acc}(E^{}_{\rm d})$ are obtained from Ref.~\cite{Abusleme:2020zyc}, where the $\alpha$'s in the the neutral radioactivity, are rejected by a PSD. Note that $ S^j_{\rm acc}(E^{}_{\rm d})$ for cosmogenic $^{11}$C should be identical to $S^i_{\rm nc}(E_{\rm d})$ for $^{11}$C induced by the $\nu_{\rm atm}$-C NC interactions. Fig.~\ref{fig:combFit} shows an example of the ML fit, where the MC data set uses \texttt{GENIE} prediction and has an exposure of 200~${\rm kt}\cdot{\rm yr}$. The upper, middle and lower panels are shown for the distributions of the time and spatial intervals between the prompt and third events, and the energy of the third event, respectively. It demonstrates that the combined fit distinguishes the NC interactions from the accidental contaminations by the cosmogenic $^{11}$C and radioactivity.
{Note that the energy distributions in the lower panel of Fig.~\ref{fig:combFit} are very useful to separate the contributions from the cosmogenic $^{11}$C and radioactivity, which can improve the accuracy of NC interaction measurements.}

\begin{figure}[!ht]
\centering
\includegraphics[width=8.5cm]{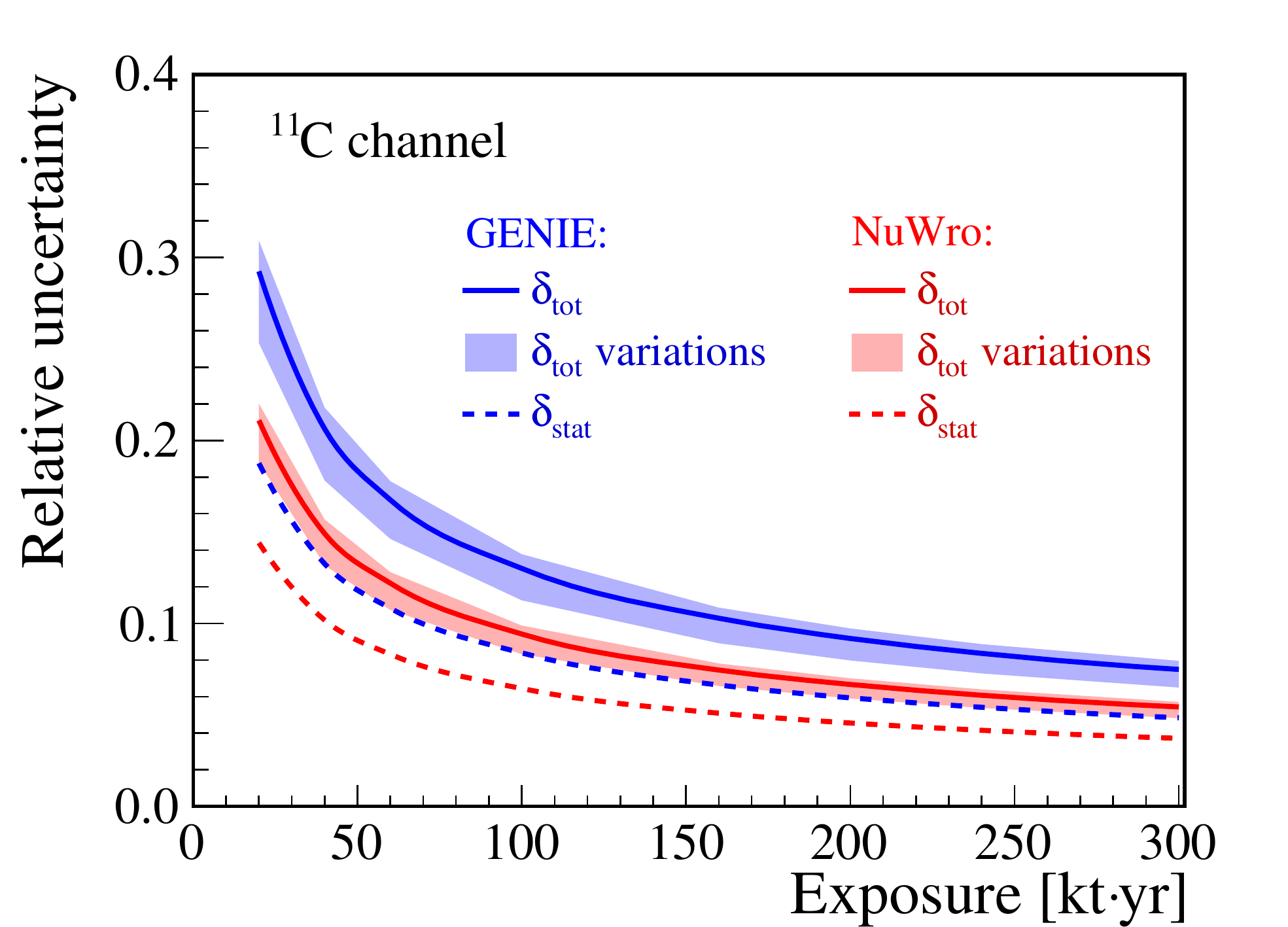}
\includegraphics[width=8.5cm]{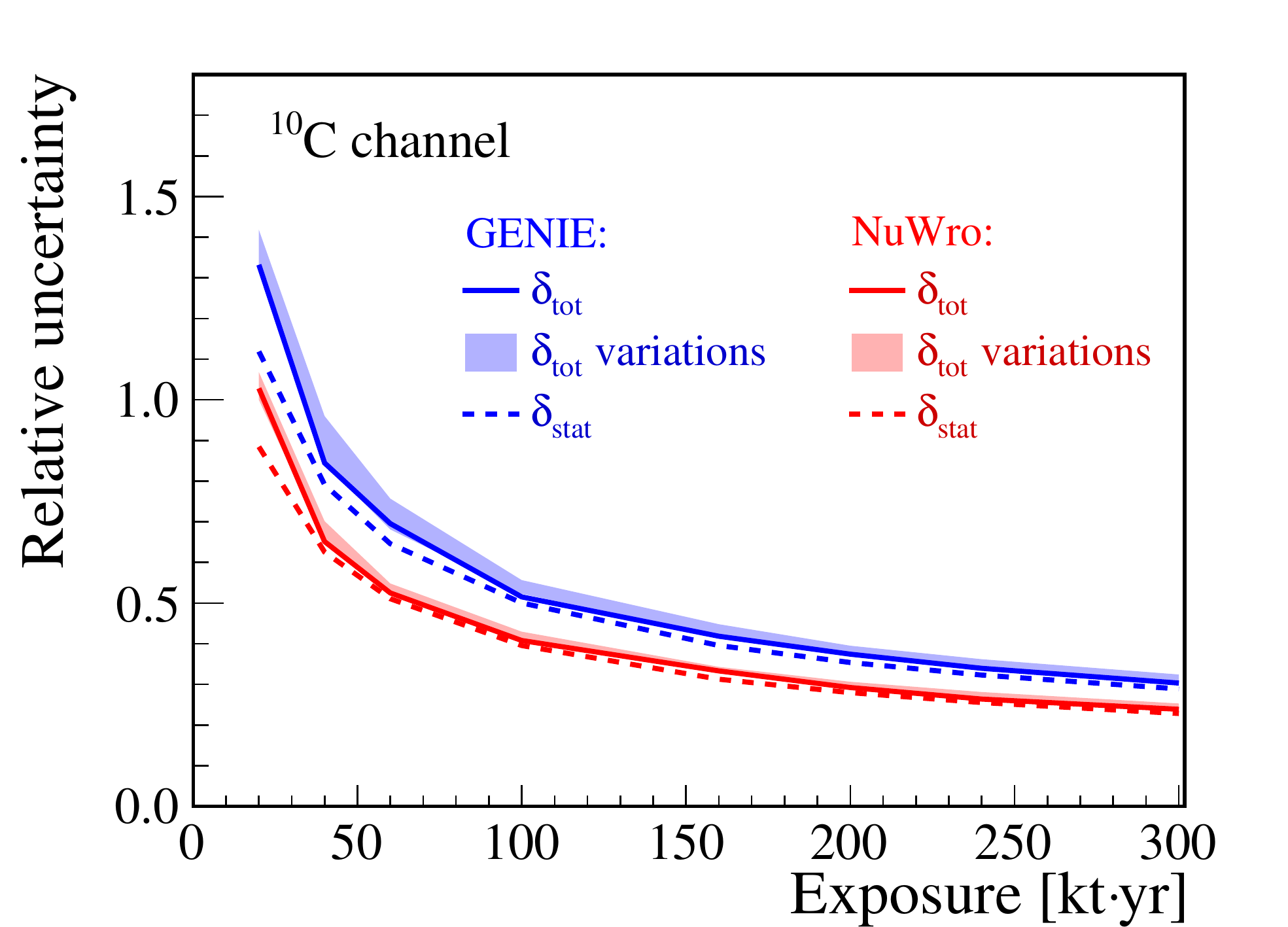}
\caption{The projected uncertainties of the {\it in situ} measurement to the two NC channels. The solid line and dashed line represent the total uncertainty $\sigma^{}_{\rm tot}$ (solid line) and statistical uncertainty $\sigma^{}_{\rm stat}$ (dashed line), respectively. The shaded bands represent the variations due to different scenarios on the LS radiopurity and the rate of the residual $^{11}$C. The blue and red colors represent the \texttt{GENIE} and \texttt{NuWro} models, respectively.}
\label{fig:unCer}
\end{figure}

 The total uncertainties ($\sigma^{}_{\rm tot}$) and statistical uncertainties ($\sigma^{}_{\rm stat}$) versus the experimental exposure are shown in Fig.~\ref{fig:unCer}. The $\sigma^{}_{\rm stat}$ is evaluated from the number of triple coincidences associated with $^{11}$C or $^{10}$C at each fixed exposure. The $\sigma^{}_{\rm tot}$ is obtained from 1000 mock data sets at each fixed exposure, where the variation in the 1000 mock data sets is due to the Poisson statistic fluctuations of the items in Table.~\ref{tab:rateMC}. Note that the  $\sigma^{}_{\rm tot}$ not only depends on the statistic fluctuations but also is affected by the accidental background level, including the neutral radioactivity and the cosmogenic $^{11}$C. We observe that the three-dimensional fit with Eq.~(\ref{eqn:pdftot}) utilizes more information and reduces the uncertainties compared to any of the two-dimensional combined fit or the one-dimensional fit.
 The $\sigma^{}_{\rm tot}$ from the fitting are significantly larger than $\sigma^{}_{\rm stat}$, in particular for the $^{11}$C channel, due to the parameters' correlations. The shaded bands in Fig.~\ref{fig:unCer} represent different scenarios on the LS radio-purity level and the rate of residual cosmogenic $^{11}$C. The upper edge accounts for the worst case that the internal radio-purity is worse by one order, while the lower edge accounts for a possibly better rejection of cosmogenic $^{11}$C. A dedicated muon simulation indicates that $^{11}$C production is accompanied by a high multiplicity of spallation neutrons. The accompanying neutrons, with a mean kinetic energy of a few tens of MeV, are captured mostly within one meter from the $^{11}$C production location. This allows us to develop a veto scheme by searching for the spallation neutrons close in space and time with respect to the third signal in a NC triple-coincidence candidate. If one coincident spallation neutron is found, the third signal is rejected. A preliminary test indicates that a rejection efficiency of 56\% for cosmogenic $^{11}$C can be achieved by a veto window of 0.35 m and 0.85 hrs, while maintaining a 97\% efficiency for the NC background. With this approach, the uncertainty curve is roughly in the middle between the solid line and the lower edge of the shaded bands. More sophisticated veto strategies can be developed for a specific detector to suppress the cosmogenic $^{11}$C further. However, it is beyond the scope of this paper.

\subsection{Implications for DSNB searches}

The above MC analysis validates the approach of measuring {\it in situ} the NC interactions associated with a suitably long-lived isotope (Category I as defined in Sec.~\ref{sec:atmNuC}). Although the analysis presented here is done for the NC interactions associated with single neutron capture only, the cases with two neutrons can as well be evaluated. Taking the correlation in Fig.~\ref{fig:rate}, one can extrapolate the number of Category II interactions from the measured Category I interactions. Eventually the $\nu^{}_{\rm atm}$ NC backgrounds can be well understood by using this data-driven approach. The outcome is to reduce the systematic uncertainty of the predicted $\nu^{}_{\rm atm}$ NC backgrounds in the searches for DSNB. A typical DSNB selection can be transformed from the criteria in Sec.~\ref{sec:InsituMeas}, by releasing the requirement of a decay-like third signal, reducing the $E^{}_{\rm p}$ range to (11, 30) MeV, and reversing the PSD cut to remove the $\nu^{}_{\rm atm}$ NC backgrounds. The residual background, $N^{}_{\rm b}$, can be estimated from the measured rates in Category I and the extrapolated rates in Category II:
\begin{equation}\label{eqn:bkg}
N^{}_{\rm b}   =  \sum_{i=^{10}{\rm C}, ^{11}{\rm C}} \frac{N^{}_i}{\varepsilon^{}_i}\cdot\left(\varepsilon_i^{\rm dsnb} + \eta\cdot\varepsilon^{\rm dsnb}_{\rm dc}\right) + N^{}_{\rm Li}\cdot\varepsilon_{\rm Li}^{\rm dsnb}
\end{equation}
$N^{}_i$ has the same definition as that in Eq.~(\ref{eqn:pdftot}), and its uncertainty is taken from Fig.~\ref{fig:unCer}. $\varepsilon^{}_i$ accounts for the corresponding total efficiency in Table.~\ref{tab:rateMC}. $\varepsilon^{\rm dsnb}_{\rm dc}$ and $\varepsilon^{\rm dsnb}_i$ represent the efficiencies of the total NC double-coincidences and the triple-coincidence channel that satisfy the DSNB criteria, respectively. $\eta$ is the extrapolation factor obtained in Sec.~\ref{sec:tripleSig}. Since the triple-coincidence associated with $^8$Li is difficult to measure, its contribution to the background is estimated entirely by simulation. $N^{}_{\rm Li}$ and $\varepsilon_{\rm Li}^{\rm dsnb}$ are the simulated number of events and the efficiency with the DSNB criteria, respectively. Note that the efficiencies $\varepsilon^{\rm dsnb}_{\rm dc}$, $\varepsilon^{\rm dsnb}_i$ and $\varepsilon^{\rm dsnb}_{\rm Li}$ include two detector-dependent efficiencies for the DSNB search, i.e., the muon veto and the PSD cut, which are assumed to be the same for the three isotopes. In a specific detector, the PSD cut efficiency might be slightly different for different isotopes. However, such an effect is not considered in this work for simplicity. Here the efficiencies in Eq.~(\ref{eqn:bkg}) are obtained from MC, and their differences between the \texttt{GENIE} and \texttt{NuWro} models are used to estimate the uncertainties of the efficiencies. According to Table.~\ref{tab:rateMC}, $\varepsilon^{}_i$ differs significantly between \texttt{GENIE} and \texttt{NuWro}, predominately due to different spectra of neutron kinetic energy shown in Fig.~\ref{fig:distrR}. However, if the prompt visible energy is greater than 7.5 MeV, the energy spectral shapes are quite similar in the $^{11}$C channel, resulting in an estimated uncertainty of $<$1\% on the ratio of $\varepsilon^{\rm dsnb}_i/\varepsilon^{}_i$. For the $^{10}$C channel, the shapes between \texttt{GENIE} and \texttt{NuWro} have a larger discrepancy, and this leads to a larger uncertainty of 5\% on the ratio of $\varepsilon^{\rm dsnb}_i/\varepsilon^{}_i$. The uncertainties on the ratios of $\varepsilon^{\rm dsnb}_{\rm dc}/\varepsilon^{}_i$ are estimated to be 4\% and 6\% for the $^{11}$C and $^{10}$C channels, respectively. The $^{8}$Li item in Eq.~(\ref{eqn:bkg}) contributes about $<1$\% to $N^{}_{\rm b}$ with a relative uncertainty of $\sim$20\%. Finally, an error propagation analysis shows that the systematic uncertainty of $N^{}_{\rm b}$ is dominated by the uncertainty of the measured $N_{\rm ^{11}C}$ shown in Fig.~\ref{fig:unCer}, which is due to the rate of the $^{10}$C channel is around 30 times less than that of the $^{11}$C channel. Even with an exposure of 40~${\rm kt}\cdot{\rm yr}$, the uncertainty estimated by the data-driven approach can surpass that estimated from the model predictions ($\sim$25\%, see Ref.~\cite{Cheng:2020aaw}). Therefore, it is promising to be constrained to a $\sim$10\% level with an exposure of 200~${\rm kt}\cdot{\rm yr}$ at a JUNO-like detector.

\section{Summary}
\label{sec:summary}

In the present work, we have developed a data-driven approach to reduce the uncertainties of the predicted NC background induced by atmospheric neutrino interactions with the $^{12}$C nuclei in LS detectors, which is expected to be of great importance for the experimental searches for DSNB. Our analysis is based on the systematic calculations in the preceding paper~\cite{Cheng:2020aaw}. In the energy range of the DSNB, the QEL process of neutrino-$^{12}$C interactions is known to be the most important. We have exploited the measurable characteristics of the QEL process in LS, such as the neutron multiplicity and the association with the suitably long-lived isotopes, which will allow us to scrutinize the nuclear models. Future large LS detectors like JUNO with numerous amounts of $^{12}$C nuclei, ultra-low radioactivity and excellent neutron tagging efficiency, are expected to make a unique contribution to the worldwide data set to improve the prediction of atmospheric neutrino NC interactions on $^{12}$C.

Taking the JUNO detector as an example, a maximum-likelihood method is developed to allow an {\it in situ} measurement of the NC interactions with a triple-coincidence signature. Furthermore, we have demonstrated that the uncertainties of NC backgrounds in the searches for DSNB can be constrained via a data-driven approach. One caveat is that, in order to significantly improve the uncertainty, the experimental collaboration needs further suppress the cosmogenic $^{11}$C and improve the LS radiopurity. Our analysis has demonstrated that with an exposure of 40~${\rm kt}\cdot{\rm yr}$ at a JUNO-like detector, the uncertainties obtained from the data-driven approach can surpass the estimated variation between models. With an exposure of 200~${\rm kt}\cdot{\rm yr}$, an uncertainty of $\sim$10\% may be achievable.

The NC background induced by atmospheric neutrinos is critical for future experimental searches for DSNB. Besides large LS detectors~\cite{An:2015jdp, Wurm:2011zn}, other large-scale detectors with advanced techniques based on water~\cite{Abe:2018uyc}, water-based LS~\cite{Askins:2019oqj} or liquid-Argon~\cite{Bueno:2007um, Acciarri:2015uup} have good potential to measure the DSNB signal. The analysis performed in the present work will be not only useful for LS detectors, but also instructive for the parallel studies of other types of detectors.

\section*{Acknowledgements}
The authors would like to thank Costas Andreopoulos, Wan-lei Guo, Bao-hua Sun and Yiyu Zhang for helpful discussions, to Guo-fu Cao and Yao-guang Wang for checking the neutron capture processes, and to Michael Wurm, Ze-yuan Yu and Shun Zhou for carefully reading the manuscript and valuable comments. This work was supported in part by National Key R$\&$D Program of China under Grant No.~2018YFA0404101, by the National Natural Science Foundation of China under Grant Nos.~11835013 and 11675203, by the Strategic Priority Research Program of the Chinese Academy of Sciences under Grant No. XDA10010100, by the China Postdoctoral Science Foundation funded project under Grant No.~2019M660793, and by the CAS Center for Excellence in Particle Physics (CCEPP).


\end{document}